\documentclass[epjST]{svjour}
\usepackage{graphics}
\usepackage{epsf,epsfig}
\def\xb{\overline{x}}

\def\als{\alpha_s}
\def\vk{{\bf k}_{\perp}}
\def\vbs{{\bf b}}

\begin{document}
\title{What can we lean about GPDs from light meson leptoproduction experiments}
\author{S.V.Goloskokov\fnmsep\thanks{\email{goloskkv@theor.jinr.ru}} }
\institute{Bogoliubov Laboratory of Theoretical  Physics,
  Joint Institute for Nuclear Research, Dubna, Russia}
\abstract{ On the basis of the handbag approach we study cross
sections and spin asymmetries for leptoproduction of various
vector and pseudoscalar mesons. Our results  are in good agrement
with high energy experiments. We analyse what information about
Generalized Parton Distributions (GPDs) can be obtained from these
reactions.
} %end of abstract
\maketitle
\section{Introduction}
\label{intro} The leptoproduction of light mesons at small
momentum transfer and large photon virtualities $Q^2$ factorizes
into a hard meson photoproduction subprocess off partons and GPDs
\cite{fact}. GPDs are complicated nonperturbative objects which
depend on 3 variables  $\xb$ -the momentum fraction of proton
carried by parton, $\xi$- skewness and $t$- momentum transfer.
GPDs contain the extensive information on the hadron structure. At
$\xi=0, t=0$ GPD become equal to the corresponding parton
distribution functions (PDFs). The form factors of hadron can be
calculated from GPDs trough the integration over $\xb$. Using Ji
sum rules \cite{ji} the parton angular momentum can be extracted.

In the light meson leptoproduction we can analyze effects of
various GPDs. The vector meson production on the unpolarized
target is sensitive to the gluon and quark GPDs $H$
\cite{gk06,gk07q}. Using GPDs $E$ we extend our analysis to the
$A_{UT}$ asymmetry for a transversally polarized target
\cite{gk08}.

The pseudoscalar meson   leptoproduction  is  analysed in
\cite{gk09,gk11}. At leading-twist these reactions are sensitive
to the GPDs $\widetilde{H}$ and $\widetilde{E}$. It was found that
essential contributions from the transversity GPDs, $H_T$ and
$\bar E_T$, are required by experiment.  Within the handbag
approach the transversity GPDs are accompanied by a twist-3 pion
wave function. It was shown that these transversity GPDs lead to a
large transverse cross section for most reactions of pseudoscalar
meson production.

Our results \cite{gk06,gk07q,gk08,gk09,gk11} on meson
electroproducion at small and moderate $x$ are in good agreement
with experimental data in the  HERA \cite{h1,zeus} HERMES
\cite{hermes} and COMPASS \cite{compass} energy range.

\section{Light meson leptoproduction in handbag approach}
\label{sec:1}

In the handbag model, the amplitude of the light meson production
off the proton  reads as a convolution of the hard partonic
subprocess
 ${\cal H}^a$ and GPDs $H^a\,(\widetilde{H}^a)$

\begin{equation}\label{amptt}
{\cal M}^{a}_{\mu'\pm,\mu +} = \, \sum_{a}\,[ \langle {H}^a
  \rangle+... ] ;\;\;\langle {H}^a\rangle \propto \sum_{\lambda}
         \int_{xi}^1 d\xb
        {\cal H}^{a}_{\mu'\lambda,\mu \lambda}(Q^2,\xb,\xi)
                                   \hat H^{a}(\xb,\xi,t)
\end{equation}
where  $a$ denotes the gluon and quark contribution with the
corresponding flavors;
 $\mu$ ($\mu'$) is the helicity of the photon (meson), and $\xb$
 is the momentum fraction of the
parton with helicity $\lambda$.

 In contrast to other analyses the subprocess amplitudes are
calculated within the modified perturbative approach (MPA)
\cite{sterman} where the quark transverse momenta $\vk$ are taken
into account together with the Sudakov suppressions. The amplitude
${\cal H}^{a}$ is a convolution of the hard part calculated
perturbatively, and the  $\vk$- dependent meson wave function
\begin{equation}\label{hsaml}
  {\cal H}^{a}_{\mu'+,\mu +}\,=
\,\frac{2\pi \als(\mu_R)}
           {\sqrt{2N_c}} \,\int_0^1 d\tau\,\int \frac{d^{\,2} \vk}{16\pi^3}
            \phi(\tau,k^2_\perp)\;
                f^{a}_{\mu',\mu}(Q,\xb,\xi,\tau,\vk).
\end{equation}
Here $\phi$ is a meson wave function, $f^a_{\mu',\mu}$ is a hard
subprocess amplitude where in the propagators  we keep the
$\vk^{\,2}$ terms. These terms are  essential under integration
near $\tau=0, \tau=1$ points.

The meson wave function is another nonperturbative object in the
model which is chosen in the simple Gaussian form
\begin{equation}\label{wave-l}
  \phi(\vk,\tau)\,\propto  a^2_M
       \, \exp{\left[-a^2_M\, \frac{\vk^{\,2}}{\tau (1-\tau)}\right]}\,.
\end{equation}
 The $a_M$ parameter determines the mean
value of the quark transverse momentum $<\vk^{\,2}>$  in the
meson. It can be seen that the wave function (\ref{wave-l})
integrated over $\vk^{\,2}$ has a form of asymptotic one $\propto
6\,\tau\, (1-\tau)$.

Together with the $<\vk^{\,2}>$ terms in the hard partonic
subprocess amplitude ${\cal H}^a$ in the MPA we consider gluonic
corrections in the form of the Sudakov factors. The Fourier
transformation of the integrals is done from the $\vk$ to $\vbs$
space where resummation and exponentiation of the Sudakov
corrections can be performed \cite{sterman}.  Details of
calculations
 can be found in \cite{gk06}.

To estimate  GPDs, we use the double distribution (DD)
representation \cite{mus99}
\begin{eqnarray}\label{ddr}
  H_i(\xb,\xi,t) =  \int_{-1}
     ^{1}\, d\beta \int_{-1+|\beta|}
     ^{1-|\beta|}\, d\alpha \delta(\beta+ \xi \, \alpha - \xb)\, f_i(\beta,\alpha,t)
\end{eqnarray}
which connects  GPDs with PDFs through the DD function $f$,
\begin{eqnarray}\label{ddf}
f_i(\beta,\alpha,t)= h_i(\beta,t)\,
                   \frac{\Gamma(2n_i+2)}{2^{2n_i+1}\,\Gamma^2(n_i+1)}\,
                   \frac{[(1-|\beta|)^2-\alpha^2]^{n_i}}
                           {(1-|\beta|)^{2n_i+1}}.
\end{eqnarray}
The functions $h$  are expressed in terms of PDFs and
parameterized as
\begin{equation}\label{pdfpar}
h(\beta,t)= N\,e^{b_0 t}\beta^{-\alpha(t)}\,(1-\beta)^{n}.
\end{equation}
Here  the $t$- dependence is considered in a Regge form and
$\alpha(t)$ is the corresponding Regge trajectory. The parameters
in (\ref{pdfpar}) are obtained from the known information about
PDFs \cite{CTEQ6} e.g, or from the nucleon form factor analysis
\cite{pauli}.

 From various meson productions at moderate HERMES and COMPASS
 energies we can get information about valence
and sea quark effects. The quarks contribute to meson production
processes in different combinations. For uncharged meson
production we have the standard GPDs. We find quark contribution
to $\rho$ production in the form: $ \propto \frac{2}{3} H^u
+\frac{1}{3} H^d$, to $\omega:\;\; \propto \frac{2}{3} H^u
-\frac{1}{3} H^d$. For production of charged and strange mesons
the transition GPDs contribute which using SU(3) symmetry can be
connected with the standard one. For example, for $\rho^+$
production the combination works: $\propto H^u -H^d$. For
pseudoscalar mesons production similar combinations of polarized
$\tilde H$ GPDs contribute. Thus, we can test various GPDs in the
different reactions.
\section{Vector meson leptoproduction}
\label{sec:2} We  apply now the results presented in section
(\ref{sec:1}) for vector meson leptoproduction on the unpolarized
target. All GPDs are modeled on the basis of the double
distribution ansatz (\ref{ddr}), (\ref{ddf}) with using the CTEQ6
\cite{CTEQ6} parameterization of PDFs.  We consider the gluon, sea
and quark GPD contribution to the amplitude. The $a_M$ parameter
in the wave function was determined from the best description of
the cross section. Some more details together with other
parameters of the model can be found in \cite{gk06,gk07q}.

This approach was found to be successful in the analysis of data
on the $\rho^0$ and $\phi$ leptoproduction \cite{gk06,gk07q}.
 In Fig.1a we show our results for the
energy dependence of the $\rho$ cross section at different $Q^2$
in the HERA energy range which describe experimental data well.

\begin{figure}[h!]
\begin{center}
\begin{tabular}{cc}
\includegraphics[width=6.1cm,height=5.3cm]{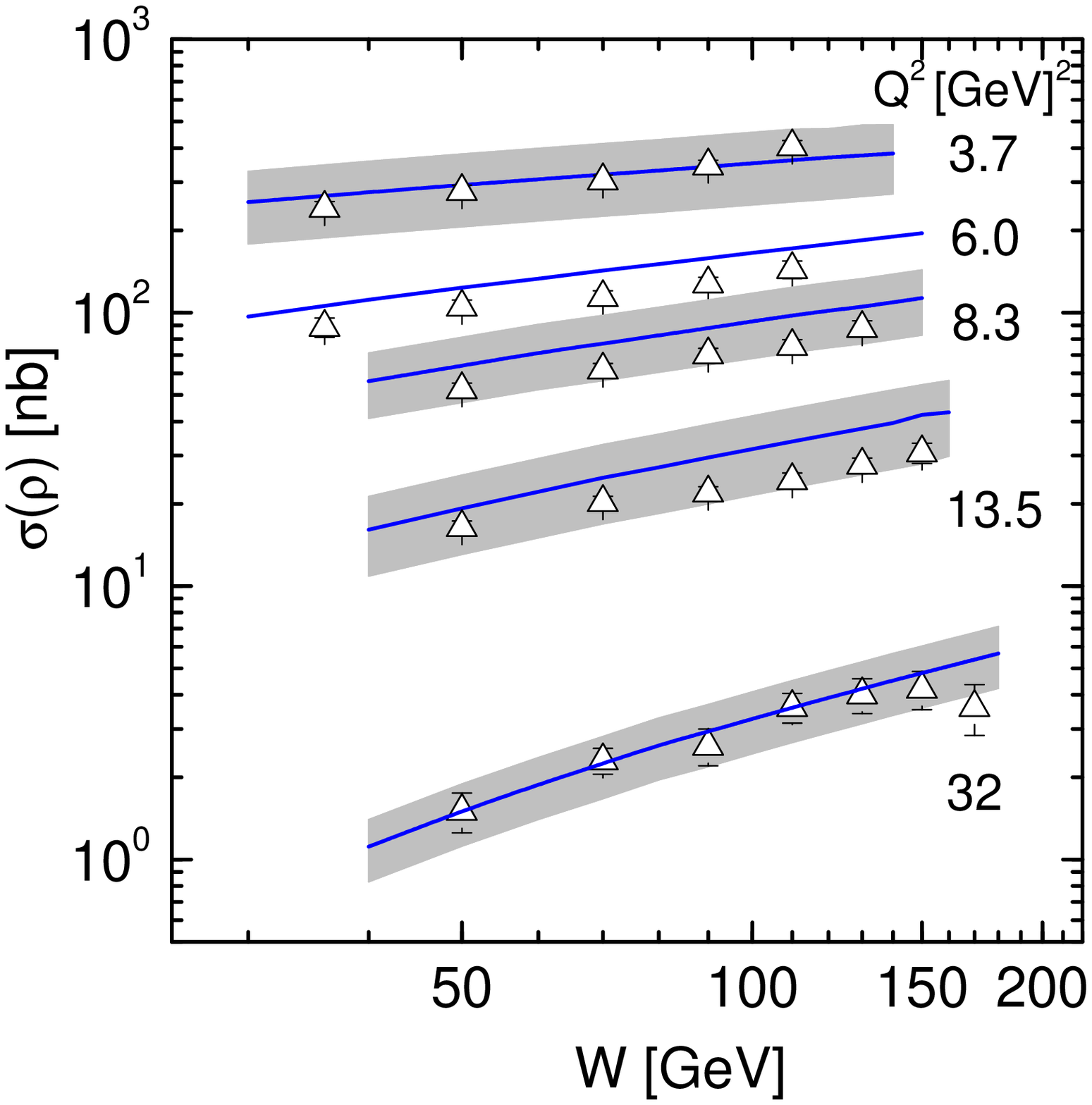}&
\includegraphics[width=6.1cm,height=5.3cm]{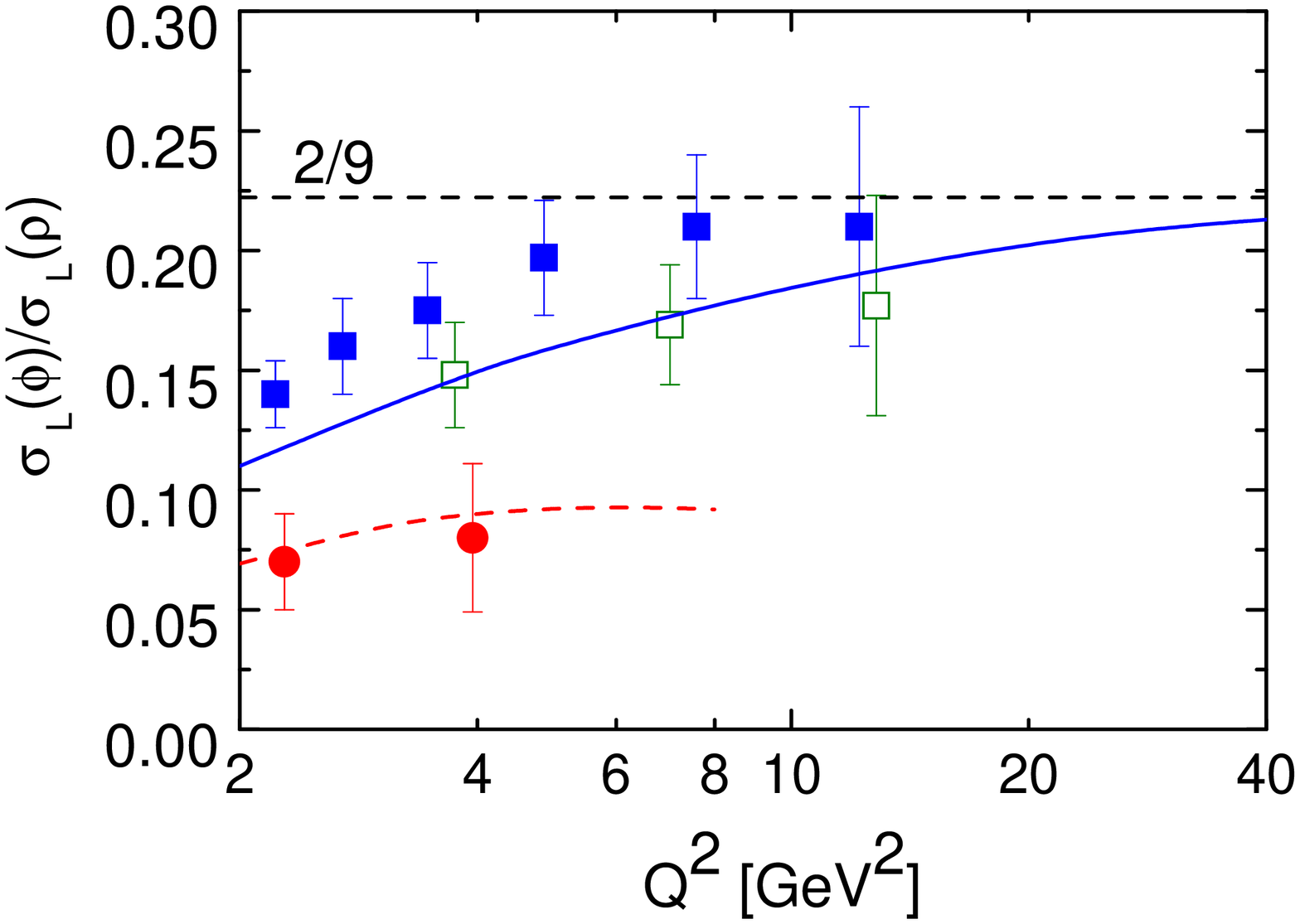}\\
{\bf(a)}& {\bf(b)}
\end{tabular}
\end{center}
\caption{{\bf(a)} The energy dependence of the $\rho$ production
cross section  at different $Q^2$ at HERA. Data: from ZEUS.
{\bf(b)} The ratio of longitudinal cross sections
$\sigma_\phi/\sigma_\rho$ at
    HERA energies- full line and HERMES- dashed line. Data are from H1  -solid, ZEUS
 -open squares, HERMES solid circles.}\label{fig:1}
\end{figure}

In Fig. 1.b the ratio of the $\phi$ and $\rho$ longitudinal cross
section is presented. If the sea GPDs are the flavor symmetric,
this ratio should be independent of $Q^2$ and be not far from 2/9.
In the model we have the flavor symmetry breaking between $\bar u$
and $\bar s$ sea
\begin{equation}\label{fsb}
H^u_{sea}=H^d_{sea}=\kappa_s\,H^s_{sea}\;\;\; \mbox{and}\;\;\;
\kappa_s=1+0.68/(1+0.52 \ln(Q^2/Q^2_0)).
\end{equation}
The symmetry breaking factor $\kappa_s$ is found from the CTEQ6M
 PDFs. Because of the flavor symmetry breaking (\ref{fsb}) the ratio of
$\sigma_\phi/\sigma_\rho$ becomes $Q^2$ dependent and very
different from the 2/9 value. The valence quark contribution to
$\sigma_\rho$ decreases this ratio at HERMES energies \cite{gk06},
Fig. 1b.

The model results for the cross section and spin observables of
electroproduced $\rho$ and $\phi$ mesons are in good agreement
with data on the unpolarized target at HERA \cite{h1,zeus},
COMPASS \cite{compass}, HERMES \cite{hermes} energies
\cite{gk06,gk07q}. Thus, we can conclude that our gluon, valence
and sea quark GPDs $H$ reproduce adequately the vector meson
leptoproduction in a wide energy and $Q^2$ range.

To study spin effects on the transversally polarized target, the
proton helicity flip amplitude is needed. It is expressed in terms
of GPD $E$
\begin{equation}
{\cal M}_{\mu' -,\mu +} \propto \frac{\sqrt{-t}}{2 m}
                          \int_{-1}^1 d\xb\,
           E^a(\xb,\xi,t)\,
           F^a_{\mu',\mu}(\xb,\xi).
\end{equation}
We  constructed the GPD $E$ from double distributions and
constrained it by the Pauli form factors of the nucleon
\cite{pauli}, positivity bounds and sum rules. The first moment of
$e^a(x)=E^a(x,0,0)$ is proportional to quark anomalous magnetic
moment
\begin{equation}
\int^1_0 dx e^a_{val}(x)=\kappa^a.
\end{equation}
The $\kappa^u$ and $\kappa^d$ factors have different signs. This
means that the GPD $E^u$ and $E^d$ should have different signs
too.  For the $\rho^0$ production, where the combination
$\frac{2}{3} E^u +\frac{1}{3} E^d$ contributes,  we  have a
compensation of quark contributions.

\begin{figure}[h!]
\begin{center}
\begin{tabular}{cc}
\includegraphics[width=6.1cm,height=5.3cm]{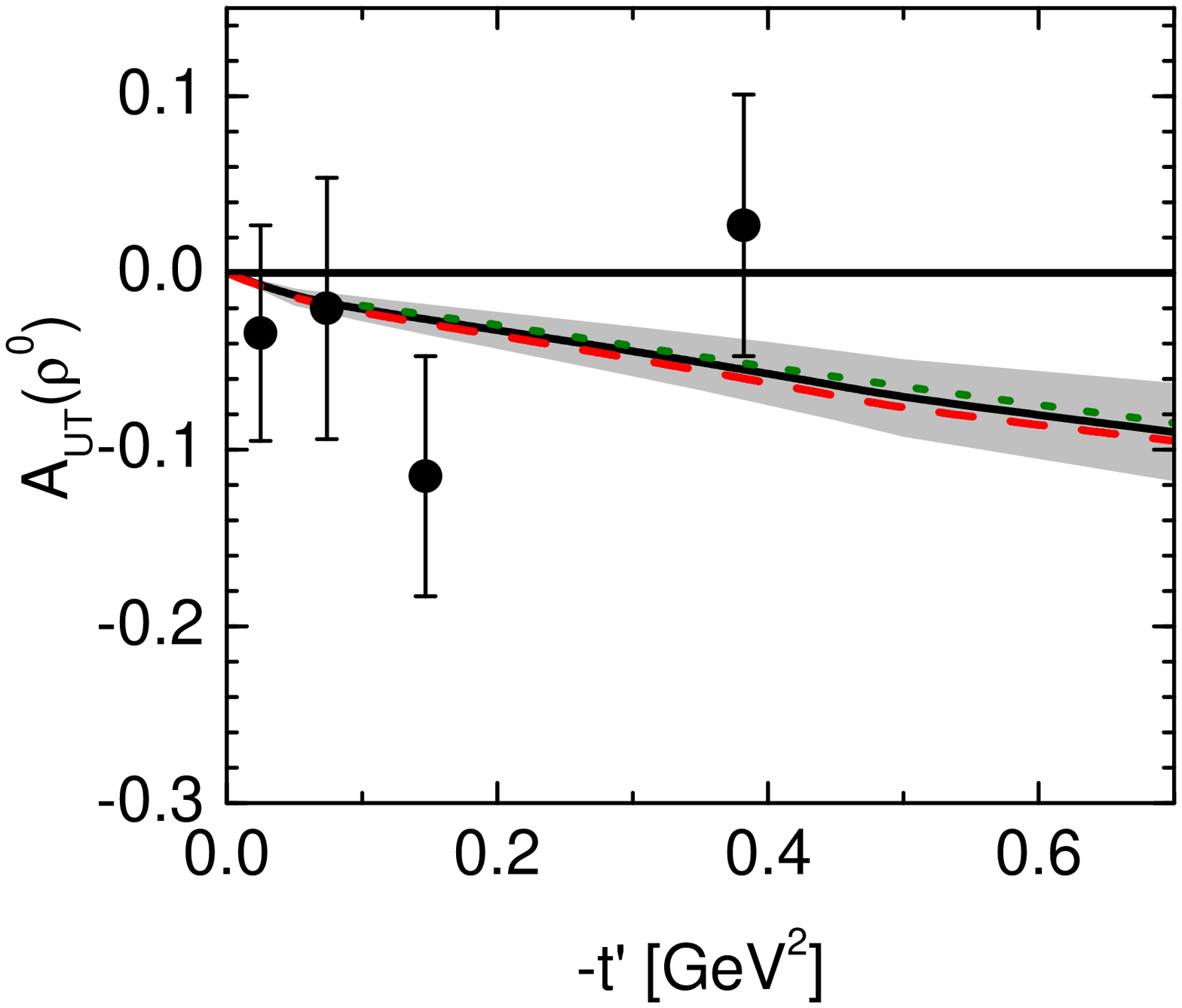}&
\includegraphics[width=6.1cm,height=5.3cm]{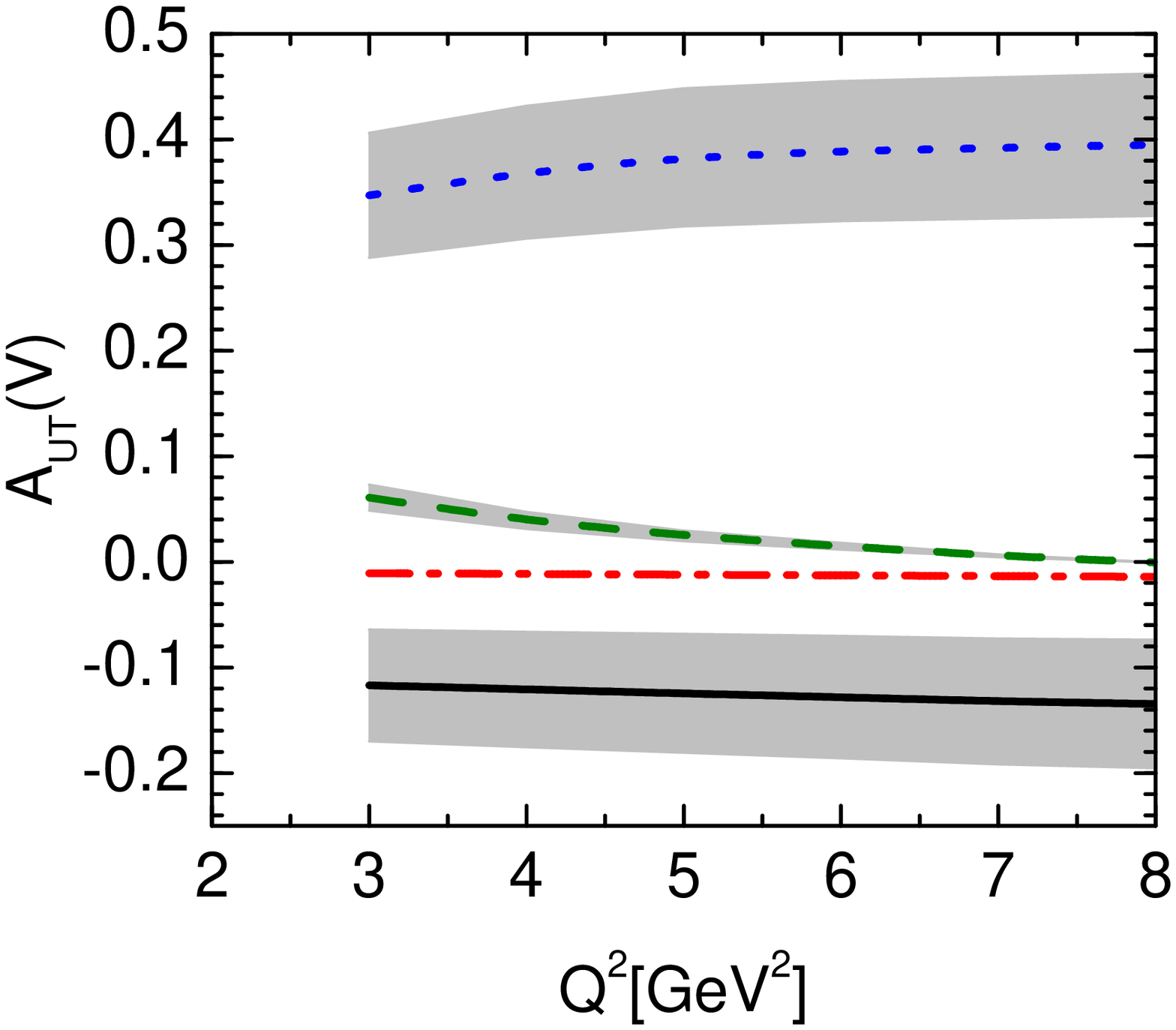}\\
{\bf(a)}& {\bf(b)}
\end{tabular}
\end{center}
\caption{{\bf(a)} Model results for the $A_{UT}$ asymmetry of the
$\rho$ production at  energy $W=5 \mbox{GeV}$ and $Q^2=3
\mbox{GeV}^2$. HERMES data are shown. {\bf(b)} Prediction for
$A_{UT}$ asymmetry for various vector meson productions at COMPASS
energy $W=10 \mbox{GeV}$. Dot-dashed line -$\rho^0$; full line
-$\omega$,
 dotted line -$\rho^+$, dashed line -$K^{* 0}$.}
\label{fig:2}
\end{figure}

The $A_{UT}$ asymmetry for transversally polarized protons  is
determined as an interference of the amplitudes connected with $E$
and $H$ GPDs.
\begin{equation}\label{aut}
A_{UT} \propto \frac{\mbox{Im}<E^*>\, <H> }{|<H>|^2}.
\end{equation}
The $H$ GPD is known from our analysis of the vector meson
leptoproduction. Our results for the $\sin(\phi-\phi_s)$ moment of
the $A_{UT}$ asymmetry of the $\rho^0$ production are shown in
Fig. 2a \cite{gk08} and describe  HERMES data \cite{hermesaut}
quite well. The $A_{UT}$ asymmetry at COMPASS is predicted to be
quite small and is in good agreement with the data \cite{sandacz}.

 Predictions for the $A_{UT}$ asymmetry at $W=5
\mbox{GeV}$ and $W=10 \mbox{GeV}$  were given in the model for the
$\omega$, $\rho^+$, $K^{*0}$ mesons \cite{gk08}. We show our
results at COMPASS energy in Fig~2.b. Our prediction for the
$\omega$ production asymmetry is negative and not small. This is
determined by enhancement of quark contributions  which are in the
$\frac{2}{3} E^u -\frac{1}{3} E^d$ combination there. Predictions
for the $\rho^+$ asymmetry is positive and rather large $\sim
0.4$. In this reaction the contribution $E^u -E^d$ works and we
have enhancement of quark contribution too. At the same time,
smallness of the $\rho^+$ cross section does not give a good
chance to measure this asymmetry. Good agreement with experimental
data at HERMES and COMPASS of the $A_{UT}$ asymmetry found in the
model shows that our estimations on GPDs $E$ are not far from
reality. However, experimental errors are quite large now and
additional experimental data for various reactions are needed to
get more information about GPDs $E$. The analysis the $A_{UT}$
asymmetry for $\omega$ at HERMES and COMPASS can help here.

\section{Leptoproduction of pseudoscalar mesons.}
Hard exclusive pseudoscalar meson leptoproduction was studied on
the basis of the handbag approach too. These reactions are
sensitive to the polarized GPDs $\tilde H$ whose parameterization
can be found in \cite{gk07q} and $\tilde E$. The pseudoscalar
meson production amplitude with longitudinally polarized photons
${\cal M}^{P}_{0\nu',0\nu}$ dominates at large $Q^2$. The
amplitudes with transversally polarized photons are suppressed as
$1/Q$. The pseudoscalar meson production amplitude can be written
as:
\begin{equation}\label{pip}
{\cal M}^{P}_{0+,0+} \propto [\langle \tilde{H}^{P}\rangle
  - \frac{2\xi mQ^2}{1-\xi^2}\frac{\rho_P}{t-m_P^2}];\;
{\cal M}^{P}_{0-,0+} \propto \frac{\sqrt{-t^\prime}}{2m}\,[ \xi
\langle \widetilde{E}^{P}\rangle + 2mQ^2\frac{\rho_P}{t-m_P^2}].
\end{equation}

The first terms in (\ref{pip}) represent the handbag contribution
to the pseudoscalar (P) meson production amplitude (\ref{amptt})
calculated within the MPA with the corresponding transition GPDs.
For the $\pi^+$ production we have the $p \to n$ transition GPD
where the combination
$\tilde{F}^{(3)}=\tilde{F}^{(u)}-\tilde{F}^{(d)}$ contributes.

The second terms in (\ref{pip}) appear for charged meson
production and are connected with the P meson pole. Here we use
the fully experimentally measured electromagnetic form factor of P
meson.

 In addition to the pion
pole and the handbag contribution which in the leading twist is
determined by the $\widetilde{H}$ and $\widetilde{E}$ GPDs a
twist-3 contribution to the amplitudes ${\cal M}_{0-,++}$, ${\cal
M}_{0+,++}$ is required by the polarized data at low $Q^2$. To
estimate this effect, we use a mechanism that consists of the
transversity GPD $H_T$, $\bar E_T$ in conjugation with the twist-3
pion wave function. For the ${\cal M}_{0-,\mu+}$ amplitude we have
\begin{equation}\label{ht}
{\cal M}^{P,twist-3}_{0-,\mu+} \propto \,
                            \int_{-1}^1 d\xb
   {\cal H}_{0-,\mu+}(\xb,...)\,[H^{P}_T+...O(\xi^2\,E^P_T)].
\end{equation}
The $H_T$ GPD is connected with transversity PDFs  as
\begin{equation}
  H^a_T(x,0,0)= \delta^a(x);\;\;\;
\delta^a(x)=C\,N^a_T\, x^{1/2}\, (1-x)\,[q_a(x)+\Delta q_a(x)].
\end{equation}
Here we parameterize the PDF $\delta$ using the model \cite{ans}.
The DD form (\ref{ddr},\ref{ddf}) is used to calculate  GPD $H_T$.

 The twist-3 contribution to the amplitude ${\cal M}_{0+,\mu+}$
 has a form \cite{gk11} similar to (\ref{ht})
\begin{equation}
{\cal M}^{P,twist-3}_{0+,\mu+} \propto \, \frac{\sqrt{-t'}}{4 m}\,
                            \int_{-1}^1 d\xb
 {\cal H}_{0-,\mu+}(\xb,...)\; \bar E^{P}_T
\end{equation}
The information on $\bar E_T$ was obtained only in the lattice QCD
\cite{lat}. The lower moments of $\bar E_T^u$  and $\bar E_T^d$
were found to be quite large, have the same sign and a similar
size. At the same time, $H_T^u$  and $H_T^d$ are different in the
sign. This means that we have an essential compensation of the
$\bar E_T$ contribution to the $\pi^+$ amplitude: $\bar
E_T^{(3)}=\bar E_T^u-\bar E_T^d$. $H_T$ does not compensate in
this process. For the $\pi^0$ production we have the opposite
case. We find here a large contribution from $\bar E_T^{\pi^0}=2/3
\bar E_T^u+ 1/3\bar E_T^d$, $H_T$ effects are not so essential
here.

\begin{figure}[h!]
\begin{center}
\begin{tabular}{cc}
\includegraphics[width=6.1cm,height=5.3cm]{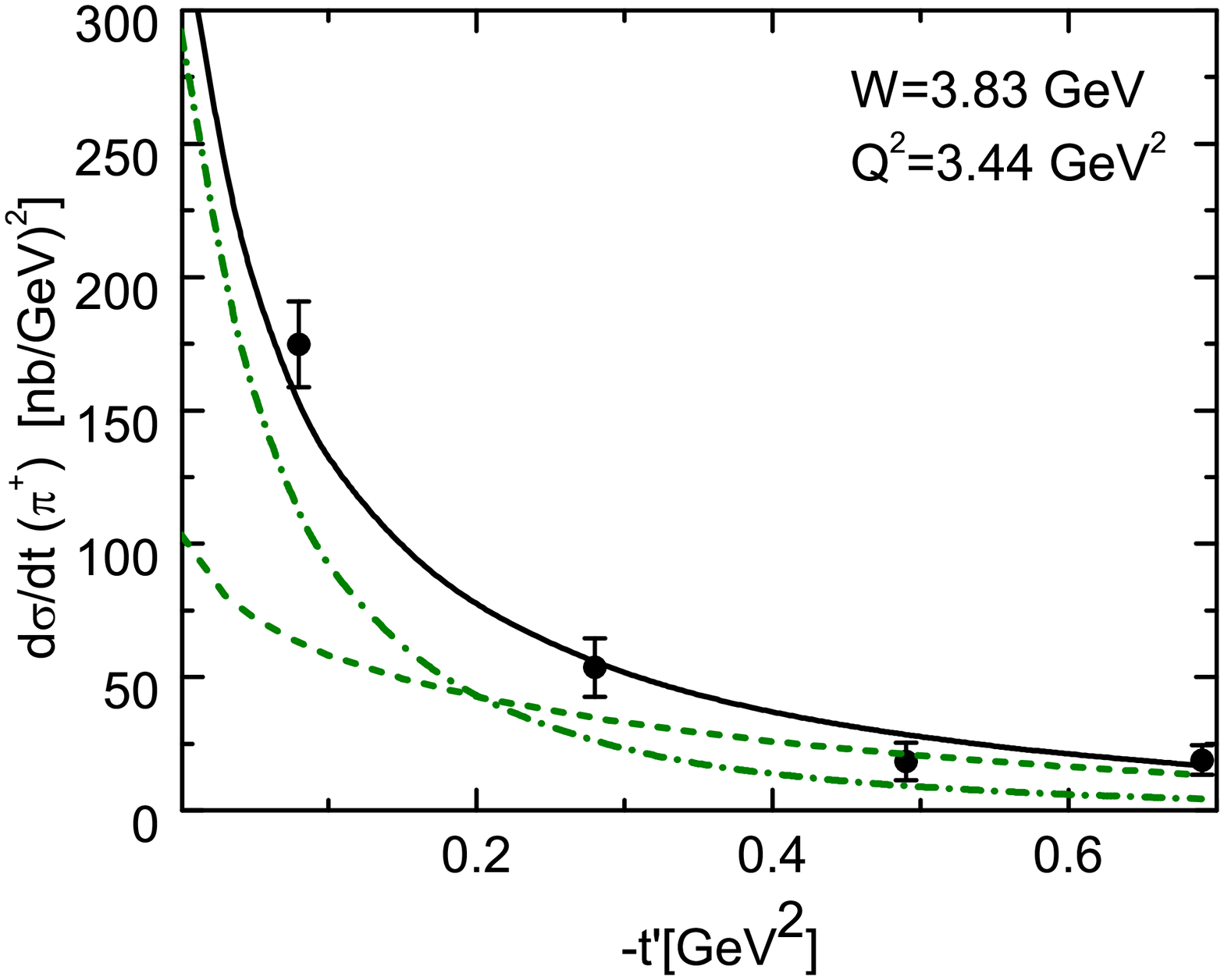}&
\includegraphics[width=6.1cm,height=5.3cm]{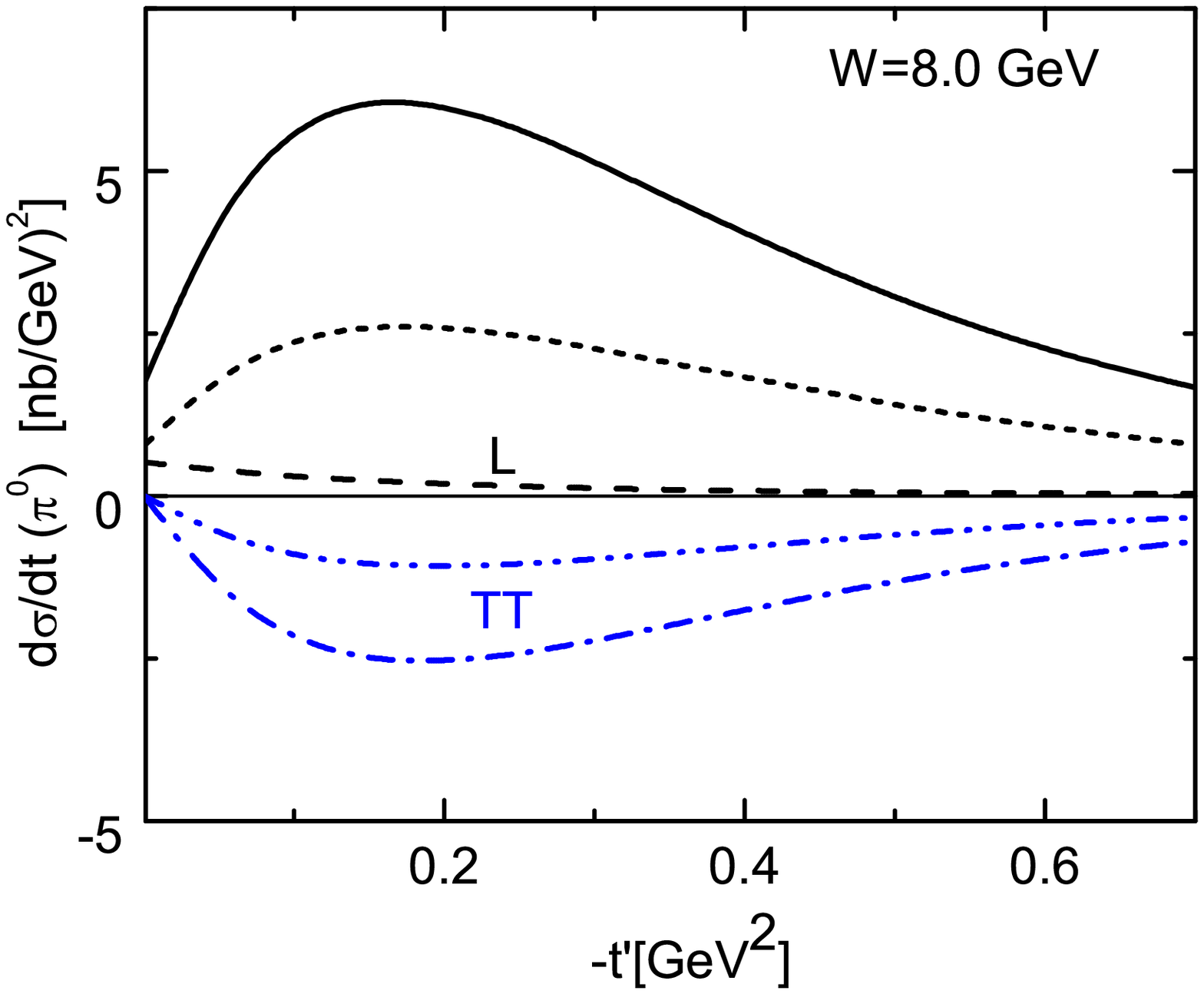}\\
{\bf(a)}& {\bf(b)}
\end{tabular}
\label{fig:3}
\end{center}
\caption{{\bf(a)}The unseparated cross section of the $\pi^+$
production -full line, together with HERMES data; dashed-dotted-
$\sigma_L$, dotted line- $\sigma_T$.    {\bf(b)} $\pi^0$
production at COMPASS. Unseparated cross section: full line-  at
$Q^2=3\mbox{GeV}^2$,  short dashed at  $Q^2=5\mbox{GeV}^2$; dashed
line- $\sigma_L$; dashed dotted- $\sigma_{TT}$ at
$Q^2=3\mbox{GeV}^2$, dashed dotted- dotted- $\sigma_{TT}$ at
$Q^2=5\mbox{GeV}^2$}
\end{figure}

In Fig. 3a, we show our results for unseparated cross section of
the $\pi^+$ production which describes fine HERMES data. The
$\sigma_L$ and $\sigma_T$ are shown as well. The longitudinal
cross section determined by leading-twist dominates at small
momentum transfer $-t < 0.2 \mbox{GeV}^2$. At larger $-t$ we find
an essential contribution from the transverse cross section.
Effects of $\bar E_T$ are small here. In Fig. 3b, we show our
results for the cross section of the $\pi_0$ production which are
quite surprising. The transverse cross section where the  $\bar
E_T$ contributions are important dominates. The longitudinal cross
section which is expected to play an essential role is much
smaller with respect to the
 transverse cross $\sigma_T$. The $\sigma_T$ cross section is determined
by the twist-3 $\bar E_T$ contributions and decreases quickly with
$Q^2$ growing, Fig. 3b.

In the same way we  calculate  the strange particle production.
The proton- hyperon transition GPDs which contribute here are
contracted by using the SU(3) flavor symmetry. For example, for
the $\gamma p \to K^+ \Lambda$ reaction we find:
\begin{equation}
  F_{p \to \Lambda} \sim -\frac{1}{\sqrt{6}} [2
F^u-F^d-F^s].
\end{equation}
Details of calculations can be found in \cite{gk11}. Our results
for the cross section of various processes are shown in Fig.4a. It
is important that we predict at HERMES for $-t>0.3 \mbox{GeV}^2$
not small and closed cross section for the $\pi^+$, $\pi^0$ and
$K^+ \Lambda$ production. This prediction is a result of large
transversity $\bar E_T$ contributions in the two last processes.
In Fig. 4b, we show our results for the beam-spin $A_{LU}$
asymmetries at HERMES for different meson channels. Predicted
asymmetries are not small except the $\pi_0$ production. Analysis
of this asymmetry at HERMES is very important because at CLAS the
$A_{LU}$ asymmetry is not small \cite{aluclas}. This analysis
gives  information about non-pole $\tilde E^{\pi^0}$ GPD.
\begin{figure}[h!]
\begin{center}
\begin{tabular}{cc}
\includegraphics[width=6.1cm,height=5.3cm]{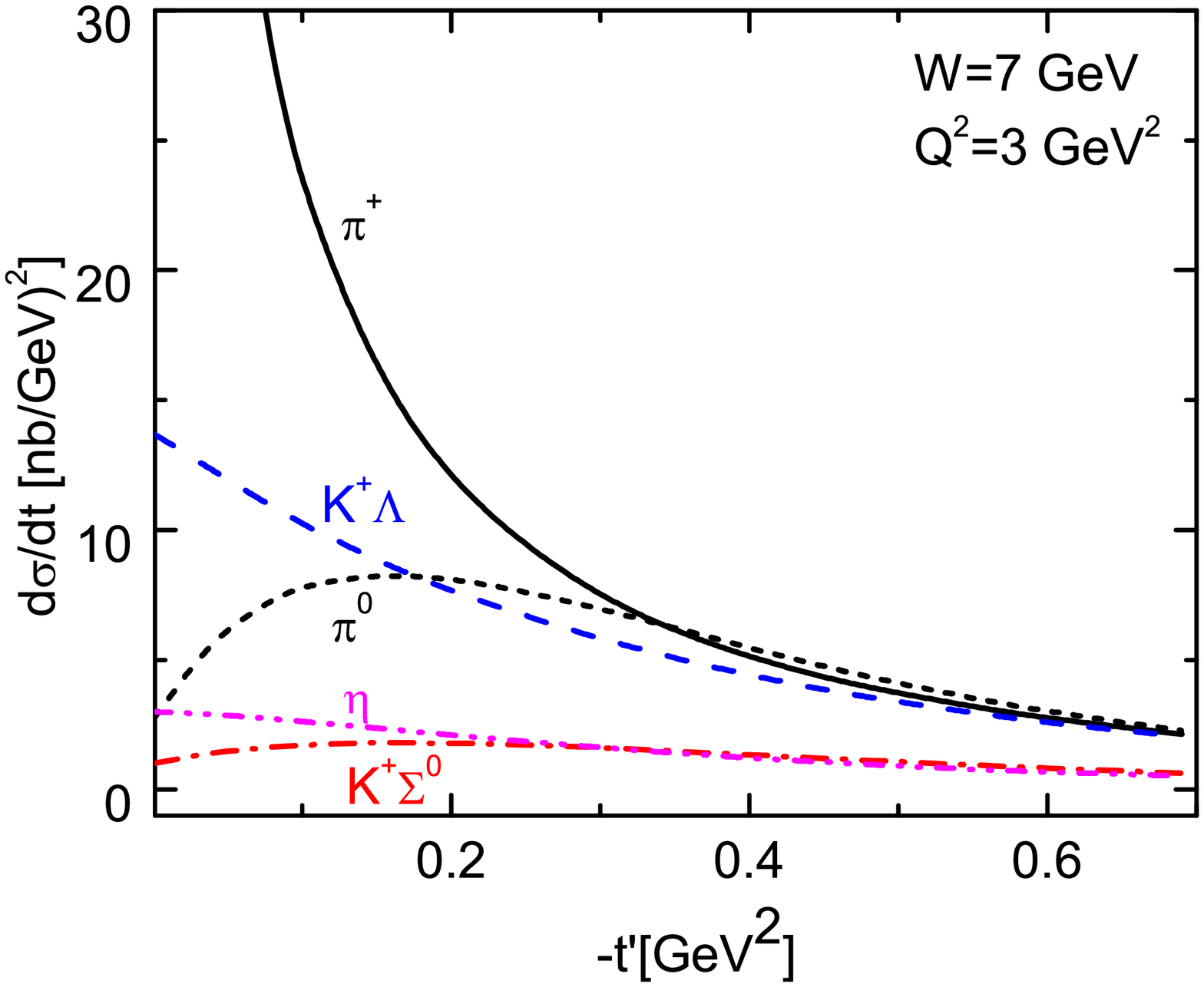}&
\includegraphics[width=6.1cm,height=5.3cm]{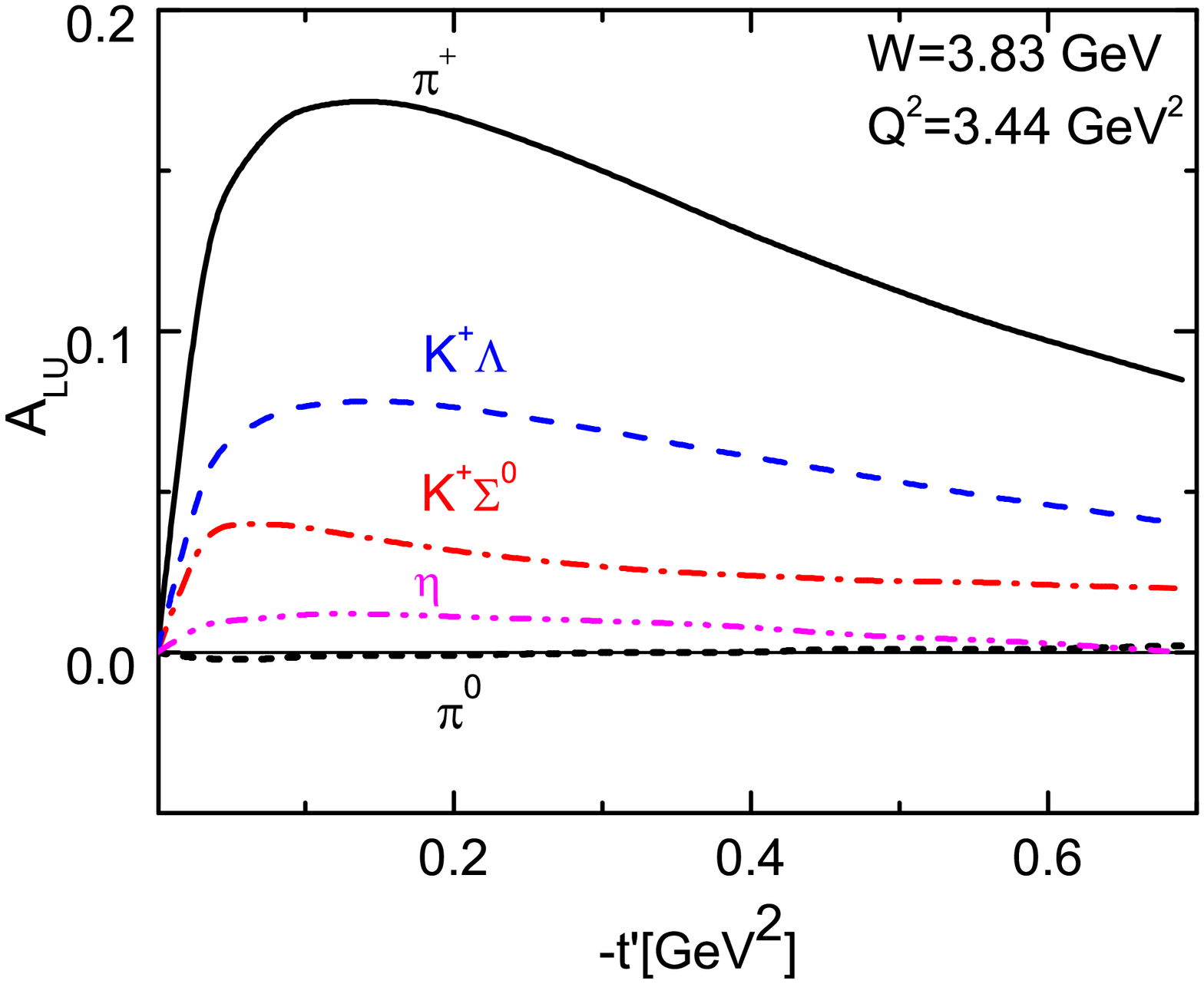}\\
{\bf(a)}& {\bf(b)}
\end{tabular}
\label{fig:4}
\end{center}
\caption{{\bf(a)}The  cross sections at COMPASS energies. {\bf(b)}
Predicted beam-spin asymmetries for at HERMES.- Both for various
  pseudoscalar meson productions}
\end{figure}

\section{Conclusion}
We  analysed  the  meson leptoproduction within the handbag
approach where the amplitude at high $Q^2$ factorizes into a hard
subprocess  and GPDs. The  hard subprocess amplitude is calculated
vising the MPA, where the transverse quark momenta and the Sudakov
factors were taken into account.

In the vector meson production on unpolarized target the gluon,
sea and valence quarks $H$ GPDs contribute which are calculated
using the CTEQ6 parameterization. Our results are in good
agreement with experiment from HERMES to HERA energies and we know
$H$ GPDs quite well.

The GPDs $E$ can be studied in experiments with a transversely
polarized target.  We model the $E$ GPDs using information on the
Pauli form factors of the nucleon and sum rules. We predict the
cross sections and $A_{UT}$ asymmetries for various meson
leptoproductions \cite{gk08}. The experimental data  available now
only for the $\rho^0$ production at HERMES and COMPASS are
described well. However, the experimental uncertainties are quite
large and  more experimental data are needed to study GPDs $E$. We
hope that analysis of the $A_{UT}$ asymmetry in the $\omega$
production at HERMES and COMPASS  can get more constraints on the
GPDs $E$.

The  amplitudes of pseudoscalar meson production in the leading
twist  are sensitive to $\tilde H$ and $\tilde E$ GPDs. However,
we have now experimental data at quite low $Q^2$. It was shown
that in this region the twist -3 effects  determined by the
transversity GPDs $H_T$ and $\bar E_T$  are very important in
understanding the cross section and spin asymmetries of the
pseudoscalar mesons production. There are some model estimations
of GPD $H_T$ \cite{ans}. For GPDs  $\bar E_T$ only the information
from the lattice is available \cite{lat}. At HERMES and COMPASS
energies the twist-3 $E_T$ effects produce a large transverse
cross section $\sigma_T$ \cite{gk11} which exceeds substantially
the leading twist longitudinal cross section. Similar behavior of
$\sigma_T$ is observed for most reactions of the pseudoscalar
meson production with the exception of channels with $\pi^+$ and
$\eta'$. This important prediction of the model can be tested
experimentally and shed light on the role of transversity effects
in these reactions.

We describe fine the cross section and spin observables for
various meson productions in a wide energy range. Thus, we can
conclude that information on GPDs discussed above should  not be
far from reality.
\bigskip

This work is supported  in part by the Russian Foundation for
Basic Research, Grant  09-02-01149  and by the Heisenberg-Landau
program.


\begin{thebibliography}{}
\bibitem{fact} X. Ji, Phys. Rev. \textbf{D55}, (1997) 7114;\\
A.V. Radyushkin,  Phys. Lett. \textbf{ B380}, (1996)  417;\\
J.C.  Collins, et al., Phys. Rev. \textbf{D56}, (1997) 2982.
\bibitem{ji} X.~Ji, Phys. Rev. Lett. \textbf{78}, (1997)  610.
\bibitem{gk06}S.V. Goloskokov, P. Kroll,
  Euro. Phys. J.  \textbf{C50}, (2007) 829.
\bibitem{gk07q} S.V. Goloskokov, P. Kroll,  Euro. Phys. J. \textbf{C53}, (2008) 367.

\bibitem{gk08} S.V.Goloskokov, P.Kroll, Euro. Phys. J. \textbf{C59}, (2009)
809.
\bibitem{gk09}  S.V.Goloskokov, P.Kroll, Euro. Phys. J.  \textbf{C65}, (2010)
137.
\bibitem{gk11}  S.V.Goloskokov, P.Kroll, Euro. Phys. J. \textbf{A47}, (2011) 112.
 \bibitem{h1} C. Adloff  et al.  [H1 Collaboration],
                        Euro. Phys. J. \textbf{C13}, (2000) 371;\\
                        C. Adloff  et al.
  F. D. Aaron et al.  [H1 Collaboration],
JHEP \textbf{1005}, (2010) 032.
\bibitem{zeus}
 S. Chekanov et al.  [ZEUS Collaboration], Nucl.Phys. \textbf{B718}, (2005)
 3;\\
  S.~Chekanov et al.  [ZEUS Collab.],  PMC Phys. \textbf{ A1} (2007) 6.

\bibitem{hermes} A. Airapetian et al. [HERMES collaboration],
     Euro. Phys. J. \textbf{C62}, (2009) 658.
\bibitem{compass}
 V.~Y.~Alexakhin
et al. [COMPASS Collab.],  Eur. Phys. J. \textbf{C52}, (2007) 255.
\bibitem{sterman} J. Botts and G. Sterman,
 Nucl. Phys. \textbf{B325}, (1989) 62.
\bibitem{mus99} I.V. Musatov and A.V. Radyushkin,
  Phys. Rev. \textbf{D61}, (2000) 074027.
  \bibitem{CTEQ6} J. Pumplin, et al.,
 JHEP \textbf{0207}, (2002) 012.
\bibitem{pauli} M. Diehl, T. Feldmann, R. Jakob and P.~Kroll,
  Eur0. Phys. J.   \textbf{C39}, (2005) 1.
\bibitem{hermesaut} A. Airapetian et al. [HERMES collaboration],
     Phys. Lett.  \textbf{B679}, (2009) 100.
\bibitem{sandacz} A.~Sandacz [COMPASS Collab.],  "Proc. of "Photon 2009", DESY,
Hamburg, 2009.
\bibitem{ans} M. Anselmino, M. Boglione, U. D'Alesio,
A. Kotzinian, F. Murgia, A. Prokudin and S. Melis,
  Nucl.  Phys.  Proc.  Suppl.   \textbf{191}, (2009) 98.
\bibitem{lat} M. Gockeler {\it et al.}
[QCDSF Collaboration and UKQCD Collaboration],
  Phys. Rev. Lett.  \textbf{98}, (2007) 222001.
\bibitem{aluclas} M. Aghasyan, H. Avakian [CLAS Collab.].
ArXiv:1103.3194 [hep-ex], (2011).
\end{thebibliography}
\end{document}